\RequirePackage{fix-cm}
\documentclass[smallextended]{svjour3}       
\smartqed  
\usepackage{graphicx}

\usepackage{verbatim}
\usepackage[utf8]{inputenc}
\usepackage[T1]{fontenc}

\usepackage{qcircuit}
\usepackage{subcaption}
\usepackage{mathtools}
\usepackage{amssymb}
\usepackage{braket}
\usepackage{xfrac}
\usepackage{graphicx}
\usepackage{graphicx, cuted}
\usepackage{braket}
\usepackage{multirow}
\usepackage{amsmath}
\usepackage{float}
\usepackage{paralist}
\usepackage{multicol}
\usepackage{caption}
\usepackage{mathrsfs}
\usepackage{verbatim}

 \usepackage{mathptmx}      
\begin{document}

\title{Obtaining entangled photons from fully mixed states using beam splitters 
}


\author{Hari krishnan S. V.         \and
        Shubhrangshu Dasgupta 
}


\institute{Hari krishnan S. V.\at
              School of Physics\\
              IISER Thiruvananthapuram. \\
              \email{hariksv9816@iisertvm.ac.in}           
           \and
            Shubhrangshu Dasgupta \at
            Department of Physics\\
            IIT Ropar, Rupnagar, Punjab\\
            India
}

\date{Received: date / Accepted: date}

\maketitle

\begin{abstract}
Preparation of entangled states of photons are useful for quantum computing and communication. In this paper, we present a simplistic protocol of entanglement generation using beam splitters with suitable reflectivity. The photons in an initial state with fully classical probability distribution pass through an optical network, made up of sequential beam splitters and are prepared in maximally entangled states. We also present the detailed theoretical analysis of entangled state generation, for an arbitrary number of photons, fed through the input ports of the beam splitters with equal probability. 

\end{abstract}

\section{Introduction}
\label{intro}

Entanglement is one of the useful resources in quantum computation and communication protocols \cite{gisin2002quantum,densecoding,teleportation}.  It is therefore important to find ways to prepare the system of interest into entangled states. The photons have been found to be most useful and relevant system, as far as the long-distance quantum communication is concerned.

In last few decades, several proposals and experiments have been put forward to generate entanglement, in either probabilistic or deterministic way. In their seminal papers, Zeilinger and his coworkers have prepared photons in polarization basis\cite{bouwmeester1997experimental} and orbital angular momentum basis\cite{mair2001entanglement}. Optical pulse passing through a nonlinear crystal gives rise to a pair of entangled photons, via parametric down conversion. Such a source of entangled photons is routinely used in quantum computation based on linear optics\cite{knill2001scheme,kok2007linear}. Though, in more recent times, efforts have been made to prepare entangled photons with higher probability using quantum dots\cite{stevenson2006semiconductor} and optical cavities\cite{garcia2008generation}, nonlinear crystals still remain the most common system in this regard, due to its availability and less stringent experimental requirement to use it.

Quantum dots\cite{michler2000quantum,ding2016demand} and nitrogen-vacancy center in Diamonds\cite{beveratos2002single} are most promising single photons sources. In this paper, we deal with the question: Can one entangle several uncorrelated single photons, using linear optical systems? It is known that the entangled photons can get uncorrelated via dephasing, in a time-scale much longer its coherence time. But to entangle several disentangled photons via a simple optical setup, one needs to employ entangling operation, namely, the beam splitters. In the following, we explicitly show how one can prepare the photons in, e.g., Bell states and NOON states, just by suitable choices of beam splitters. We emphasize that in our case, the entanglement is in Fock state basis, not in polarization basis.

Our protocol is different from that used for entanglement purification \cite{PhysRevLett.69.2881,PhysRevLett.76.722,PhysRevLett.77.2818} and entanglement concentration/distillation \cite{Nielsen,concentration}. Entanglement purification is the process of extracting a smaller number of maximally entangled pairs out of a large number of less-entangled pairs using only local operations and classical communication. On the other hand, entanglement concentration involves obtaining a maximally entangled state from many available copies of a partially entangled state. Both these strategies however, require some amount entanglement to begin with. Here we propose a method to achieve entanglement using only maximally mixed states as input. This therefore becomes quite unique in the sense that one obtains purely quantum state, from a state with classical probability distribution, that too with a simple optical network with suitably chosen beam splitters. We emphasize that our protocol, though probabilistic, does not require any measurement and the states thus prepared are ready for further network use. The states thus prepared can be verified using standard tomographic measurement.

The structure of the paper is as follows. In Sec. II, we will analyze the basic tools of calculation and estimate the probabilities of obtaining entangled states for a specific example. In Sec. III, we will present a detailed theoretical result of evolution of a fully mixed states through a generalized optical network. We conclude the paper in Sec. IV.

 \section{Entanglement extraction using beam splitters}
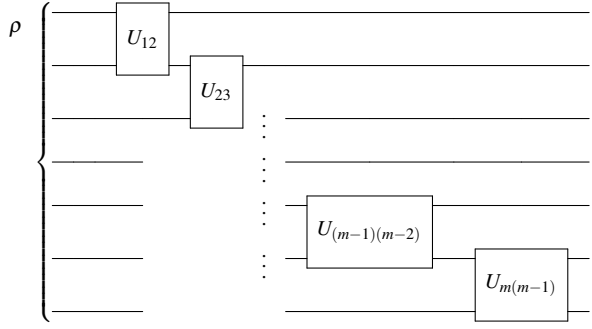
\begin{figure}[ht]

\leavevmode
\centering
\Qcircuit @C=1em @R=1.6em {
\lstick{} & \qw & \qw & \multigate{1}{U_{12}} &\qw&\qw& \qw& \qw& \qw& \qw&\qw\\
\lstick{} & \qw & \qw& \ghost{U_{12}}& \multigate{1}{U_{23}}& \qw&\qw& \qw& \qw& \qw&\qw\\
\lstick{} & \qw & \qw&\qw&\ghost{U_{23}}&  \vdots &&\qw&\qw&\qw&\qw \\
\lstick{} & \qw&\qw & \qw& &\vdots&&\qw&\qw&\qw&\qw\\
\lstick{} & \qw & \qw& \qw&&\vdots&& \multigate{1}{U_{(m-1)(m-2)}}& \qw& \qw& \qw\\
\lstick{} & \qw & \qw&\qw&&\vdots&&\ghost{U_{(m-1)(m-2)}}& \qw& \multigate{1}{U_{m(m-1)}}& \qw\\
\lstick{} & \qw &\qw& \qw& &&&\qw&\qw&\ghost{U_{m(m-1)}}& \qw
\inputgroupv{1}{7}{.8em}{.8em}{\rho}
}
  \caption{Beam splitter arrangement. The state $\rho$, as in equation (\ref{rhomix}) is the mixed state used as input. All the beam splitters have the same, but arbitrary values of TR.}
\label{fig:1}
\end{figure}

 Consider an $m$-channel interferometer into which $n$ photons are provided as input ($n\leq m$) \cite{wang2017high} (see figure \ref{fig:1}). We shall restrict our attention to states of the form:
 \begin{equation}
   \ket{i_1i_2 \dots \mbox{m times..}i_m} = (a_1 ^\dagger)^{i_1} (a_2 ^\dagger)^{i_2} \dots (a_m ^\dagger)^{i_m}\ket{00\dots \mbox{m times..}0}
 \label{eqn:somelabel}
 \end{equation}
 with $i_1,i_2 \dots i_m$ $\in$ $\{0,1\}$ and $\sum_{j=1}^{m}i_j = n $. Here $a_j^\dagger$ refers to the creation operator for the $j$-th optical mode of the photons.

 Using the states in equation \ref{eqn:somelabel}, one can consider the mixed state, $\rho$, as the input to the interferometer and as given by
 \begin{equation}
   \rho = \dfrac{1}{^mC_n}\left[\sum_{\substack{i_1,i_2 \dots i_m \in \{0,1\}\\ \sum_{j=1}^{m}i_j = n }}  \ket{i_1i_2 \dots \mbox{m times..}i_m}\bra{i_1i_2 \dots \mbox{m times..}i_m}\right]
   \label{rhomix}
 \end{equation}
 This state evolves unitarily through an optical network which is made up of two-input beam splitters arranged sequentially with arbitrary transmittance and reflectance (TR) (see figure \ref{fig:1}). Our aim is to find for which TR ratio, the probability of obtaining maximally entangled states at two output ports becomes maximum. It is important to note that the state (\ref{rhomix}) corresponds to a physical situation, when $n$  photons are fed into $m$ input ports of the interferometer, all with equal probability $1/{^mC_n}$, where $C$ denotes combination.

In figure \ref{fig:1}, we have used several beam splitters in a sequential manner. The unitary operator corresponding to a beam splitter is given by $U=\exp(i\theta (a^\dag b +b^\dag a)$ for two input modes $a$ and $b$. Such an operation leads to entangling the two modes at the output, as described in the following matrix relation:
\begin{equation}
    \begin{pmatrix}
a_i ^\dagger \\
a_j ^\dagger
\end{pmatrix}
\rightarrow
\begin{pmatrix}
\cos \theta & i \sin \theta\\
i \sin\theta & \cos \theta
\end{pmatrix}
    \begin{pmatrix}
a_i ^\dagger \\
a_j ^\dagger
\end{pmatrix}\;,
\end{equation}
where $\cos \theta $ and $i \sin \theta $ are the transmission and reflection coefficients, respectively. The factor $i$ arises due to phase change upon reflection.

We apply these operations sequentially for all the beam splitters (see figure \ref{fig:1}). Finally, the output of all the $m$ ports is traced over all but the first two channels. We then find out the suitable values of $\theta$ which maximises the probability of obtaining Bell and NOON states, at the first two ports. Note that the taking trace over the rest $(m-2)$ output ports essentially means that we do not need to make any measurement in those ports. The photons in those ports can simply be discarded or reused for other purposes.

We interrogate the probabilities of preparing the following states at the first two output ports:

(1) The single photon maximally entangled states (Bell states):
\begin{align}
    \ket{\psi}_{\pm} &= \frac{1}{\sqrt{2}} \left( \ket{10} \pm \ket{01} \right) \;\;\;  \text{and} \\
     \nonumber  \ket{\phi}_{\pm} &= \frac{1}{\sqrt{2}} \left( \ket{00} \pm \ket{11} \right)
\end{align}

(2) NOON states:
\begin{equation}
    \ket{\psi}_{\text{NOON}}^\pm = \frac{1}{\sqrt{2}} \left( \ket{\text{n}0} \pm \ket{0\text{n}} \right)\;.
\end{equation}
Note that the NOON states are useful for precision measurement of phase \cite{giovannetti2006quantum} in the context of quantum metrology \cite{dowling2008quantum}.


 \subsection{A simple example}
Before extending the problem into generalized case, let us start with $m=3$ and $n=2$, i.e., 2 photons are fed into 3 input ports with equal probabilities. The input state can then be written as
\begin{equation}
    \rho =\frac{1}{3}[\ket{101}\bra{101} +
\ket{110}\bra{110} + \ket{011}\bra{011}]\;,
\label{input1}
\end{equation}
where $\ket{101}$ = $ a_1^\dagger a_3 ^\dagger \ket{000}$, $\ket{110}$ = $ a_1^\dagger a_2 ^\dagger \ket{000}$, and 
$\ket{011}$ = $ a_2^\dagger a_3 ^\dagger \ket{000}$.

The beam splitter transforms the following states as:
\begin{align}
\ket{10} = \hspace{8pt} a_1 ^\dagger \ket{00} & \rightarrow (a_1 ^\dagger \cos \theta + a_2 ^\dagger i \sin \theta) \ket{00} \label{basiceqn1}\\[3pt] \nonumber &= \cos  \theta \ket{10} + i \sin\theta \ket{01}\\[3pt]
\ket{01} = \hspace{8pt} a_2 ^\dagger \ket{00} &\rightarrow (a_1 ^\dagger i \sin \theta + a_2 ^\dagger \cos \theta) \ket{00} \label{basiceqn2}\\[3pt] \nonumber &= i \sin \theta \ket{10} + \cos \theta \ket{01}\\[3pt]
\ket{11} = a_1 ^\dagger a_2 ^\dagger \ket{00} &\rightarrow (a_1 ^\dagger \cos \theta + a_2 ^\dagger i \sin \theta)(a_2 ^\dagger \cos \theta + a_1 ^\dagger i \sin \theta)\ket{00} \label{basiceqn3}\\[3pt] \nonumber   &= (\ket{20} + \ket{02})i \frac{\sin 2\theta}{\sqrt{2}} +  \cos 2\theta \ket{11}\\[3pt]
\ket{20} = \frac{a_1 ^\dagger a_1 ^\dagger}{\sqrt{2}} \ket{00} &\rightarrow\frac{1}{\sqrt{2}} (a_1 ^\dagger \cos \theta + a_2 ^\dagger i \sin \theta) (a_1 ^\dagger \cos \theta + a_2 ^\dagger i \sin \theta) \ket{00} \label{basiceqn4} \\
\nonumber& = \cos^2 \theta \ket{20}  -  \sin^2 \theta \ket{02} +  i \frac{\sin 2\theta \ket{11}}{\sqrt{2}}
\end{align}

 The optical network under consideration has two beam splitters corresponding to the respective unitary transformations, $U_{12}$ and $U_{23}$, where $U_{ij}$ operates on the input modes $   \begin{pmatrix}
a_i ^\dagger \\
a_j ^\dagger
\end{pmatrix} $. It is clear from the figure \ref{fig:1} that one of the output ports of the first beam splitter acts as one of the input ports for the second. 

Using the equations (\ref{basiceqn1}-\ref{basiceqn4}), we find that the state $\ket{101}$ evolves through the network in the following way:
\begin{align*}
  U_{12}\ket{101} &= \cos  \theta \ket{101} + i \sin\theta \ket{011}\\
\   U_{23}U_{12}\ket{101} &= \cos \theta \left(i \sin \theta \ket{110} + \cos \theta \ket{101}\right) + i \sin\theta \left[(\ket{020} + \ket{002}) i \frac{\sin 2\theta}{\sqrt{2}} +  \cos 2\theta \ket{011}\right]\\
    & =  i \cos \theta \sin \theta \ket{110} + \cos^2 \theta \ket{101} + i \sin\theta \cos 2\theta \ket{011}  - \sin\theta \bigg[(\ket{020} + \ket{002}) \frac{\sin 2\theta}{\sqrt{2}} \bigg]
    \end{align*}
Proceeding in a similar fashion, we get;
\begin{align*}
    U_{23}U_{12}\ket{110} & =  \ket{200} i \frac{\sin 2\theta}{\sqrt{2}} +  i \frac{\sin 2\theta}{\sqrt{2}} \cos^2 \theta \ket{020}  - i \frac{\sin 2\theta}{\sqrt{2}} \sin^2 \theta \ket{002} + \frac{\sin^2 2\theta}{2} \ket{011} \\ \nonumber & \quad +  \cos 2\theta \cos  \theta \ket{110}  + i  \cos 2\theta \sin\theta \ket{101}\\
     U_{23}U_{12}\ket{011}
    & =  - \sin^2 \theta \ket{110} + i\sin \theta \cos \theta \ket{101} + \cos \theta (\ket{020} + \ket{002}) \frac{i \sin 2\theta}{\sqrt{2}} \\ \nonumber & \quad   + \cos \theta \cos 2\theta \ket{011}
\end{align*}

Thus the input state (\ref{input1}) gets transformed as
\begin{align*}
    \rho \rightarrow \rho_{out}  = \frac{1}{3}&\big[ \ket{\psi_1}\bra{\psi_1}+\ket{\psi_2}\bra{\psi_2}+\ket{\psi_3}\bra{\psi_3} \big]\;,
\end{align*}
where $\ket{\psi_1} = U_{23}U_{12}\ket{101}$, $\ket{\psi_2} = U_{23}U_{12}\ket{110}$, $\ket{\psi_3} = U_{23}U_{12}\ket{011}$.

\begin{figure}
     \centering
     \begin{subfigure}[b]{0.4\textwidth}
         \centering
         \includegraphics[width=1\textwidth]{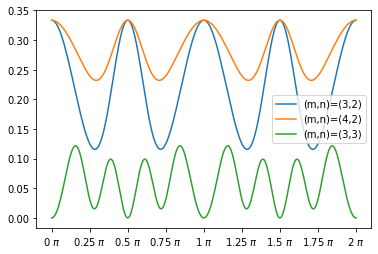}
         \caption{} 
         \label{fig:three sin x}
     \end{subfigure}
     \hfill
     \begin{subfigure}[b]{0.4\textwidth}
         \centering
         \includegraphics[width=1\textwidth]{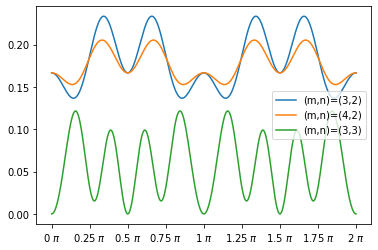}
         \caption{}
         \label{fig:five}
     \end{subfigure}
     \hfill
     \begin{subfigure}[b]{0.4\textwidth}
         \centering
         \includegraphics[width=1\textwidth]{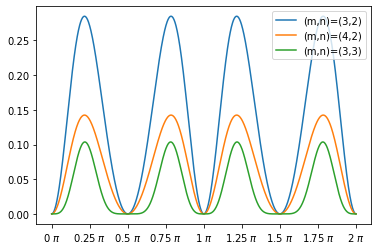}
         \caption{}
         \label{fig:five over}
     \end{subfigure}
     \hfill
     \begin{subfigure}[b]{0.4\textwidth}
         \centering
         \includegraphics[width=1\textwidth]{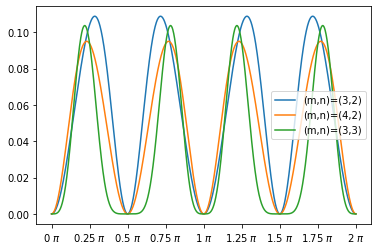}
         \caption{}
         \label{fig:five over x}
     \end{subfigure}
        \caption{Variation of the probability of obtaining the states (a) $\ket{\psi}_{\pm}$, (b) $\ket{\phi}_{\pm}$, (c) $\ket{\psi}_{NOON}^+$ and (d) $\ket{\psi}_{NOON}^-$ at the first two output ports, with $\theta$, for different values of $(m,n)$. In (a), for $(m,n)=(3,2)$, the probability becomes maximum at $\theta=r\pi/2$ and minimum at $\theta=(2r+1)\pi/4$ ($r\ge 0$ is an integer). In (b), for $(m,n)=(3,2)$, the maximum occurs at $\theta=r\pi/6$ ($r=2,4,8,10$ etc.). In (c), the probability becomes maximum at odd multiples of $\theta=\pi/4$. It is clear from (d) that the probability of obtaining the $\ket{\psi}_{NOON}^-$ is at the most $\approx 10$\% for suitable choices of $\theta$.}
        \label{fig: graphs}
\end{figure}

Next, the state $\rho_{out}$ can be partially traced over the third channel. We show in figures \ref{fig: graphs} the fidelity of obtaining Bell and NOON states as a function of $\theta$. We also compare the similar variations for  $m = n = 3$ and $m = 4, n = 2$. It is clear that for $(m,n)=(3,2)$, the probability of obtaining $|\psi_\pm\rangle$ states at the first two output ports becomes maximum, when $\theta$ becomes even multiple of $\pi/4$. It is known that for $\theta=\pi/2$ (such that the transmission coefficient $\cos\theta$ vanishes), the beam splitter works as a full reflector. In such a case, fields at both the input ports get reflected into the output ports without any transmission loss, thereby maximizing the probability. On the other hand, for a 50/50 beam splitter ($\theta$ is an odd multiple of $\pi/4$), the probability gets minimized. When the number of input ports is increased to 4, the probability at $\theta = (2r+1)\pi/4$ (for a 50/50 beam splitter; $r\ge 0$ is an integer)improves further. However, for $m=n=3$, the probability of obtaining the Bell states $|\psi_\pm\rangle$ becomes much less for all values of $\theta$. Similar results can be seen for the state $\ket{\psi}_{NOON}^+$.

\begin{figure}
     \centering
     \begin{subfigure}[b]{0.4\textwidth}
         \centering
         \includegraphics[width=1\textwidth]{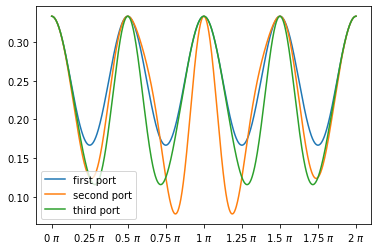}
         \caption{}
         \label{fig:psi+}
     \end{subfigure}
     \hfill
          \begin{subfigure}[b]{0.4\textwidth}
         \centering
         \includegraphics[width=1\textwidth]{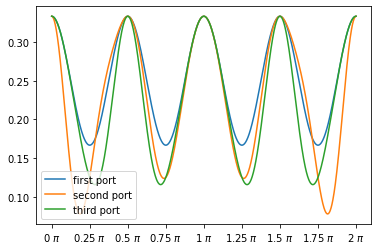}
         \caption{}
         \label{fig:psi-}
     \end{subfigure}
     \hfill
     \begin{subfigure}[b]{0.4\textwidth}
         \centering
         \includegraphics[width=1\textwidth]{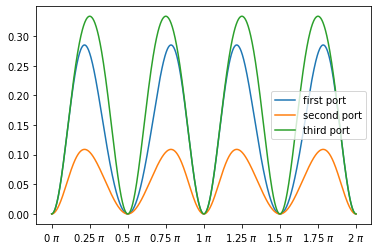}
         \caption{}
         \label{fig:noon+}
     \end{subfigure}
     \hfill
     \begin{subfigure}[b]{0.4\textwidth}
         \centering
         \includegraphics[width=1\textwidth]{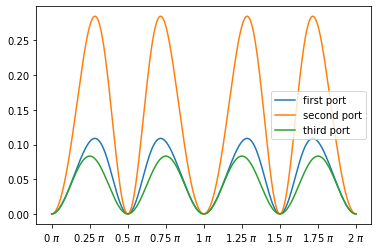}
         \caption{}
         \label{fig:noon-}
     \end{subfigure}
   \hfill
     \begin{subfigure}[b]{0.4\textwidth}
         \centering
         \includegraphics[width=1\textwidth]{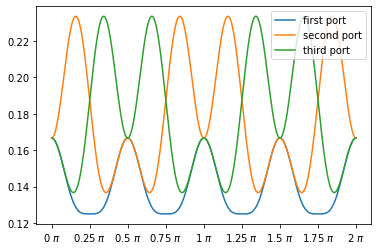}
         \caption{}
         \label{fig:phi}
     \end{subfigure}
        \caption{Variation of the probability of obtaining the states (a) $\ket{\psi}_{+}$, (b) $\ket{\phi}_{-}$, (c) $\ket{\psi}_{NOON}^+$ and (d) $\ket{\psi}_{NOON}^-$, and (e) $\ket{\psi}_{\pm}$  at different pairs of output ports, with $\theta$, for $(m,n)=(3,2)$. The reference to port the legend boxes refer to the port, upon which the partial trace is taken. In (a) and (b), the probability becomes maximum at $\theta=r\pi/2$ ($r\ge 0$ is an integer) in al the pairs of ports. In (c), the maximum occurs at $\theta=(2r+1)\pi/4$ ($r\ge 0$ is an integer) in the first two ports, whereas as in (d), the probability of obtaining $\ket{\psi}_{NOON}^-$ state becomes maximum in the first and the third ports. The probabilities of obtaining the Bell states  $\ket{\phi}_{\pm}$ are however maximum at either first two ports or in the first and the third ports, for suitable choices of $\theta$.}
        \label{fig: port graphs}
\end{figure}

If the probability of obtaining $\ket{\psi_\pm}$ across the first and the third ports, instead of the first two output ports, is calculated, it is found to exhibit maximum at $\theta=2r\pi$ ($r$ is an integer) (see figure \ref{fig: port graphs}).


\section{Extension to general m,n}
 Now the focus can be shifted onto calculating the probabilities for general $m$ and $n$. Given an input state $\rho$, the probabilities of pure  entangled states can be found easily once we figure out U$\rho$. We approach this problem of finding U$\rho$, first with the case where $m=n$, and then extend it to general case of $m\neq n$.

\subsection{For m=n.}
 In this situation, the input is $\ket{11...\text{m times}...11}\bra{11...\text{m times}...11}$. This input can be rewritten as $\ket{\psi}\bra{\psi}$ with $\ket{\psi} = \prod_{i=1}^{m}a_i\ket{00...\text{m times}...00}$.
 Applying the unitary sequences on this state, each creation operator transforms as:
  \begin{align*}
        & a_1^\dagger \rightarrow a_1^\dagger \textquotesingle =  a_1^\dagger \cos\theta + i \sin\theta[a_2^\dagger \cos \theta + i \sin \theta [a_3^\dagger \cos \theta + ....i \sin\theta[a_{m-1}^\dagger \cos \theta + a_m^\dagger i \sin \theta]]] \\[5 pt]
        & a_2^\dagger \rightarrow a_2^\dagger \textquotesingle =  a_1^\dagger i \sin \theta + \cos \theta[ a_2^\dagger \cos \theta + i \sin \theta[ a_3^\dagger \cos \theta +... i \sin\theta[a_{m-1}^\dagger \cos \theta +a_m^\dagger i \sin \theta]] \\[5 pt]
       & a_3^\dagger \rightarrow a_3^\dagger \textquotesingle = a_2^\dagger i \sin \theta + \cos \theta[ a_3^\dagger \cos \theta + i \sin \theta[ a_4^\dagger \cos \theta +... i \sin\theta[a_{m-1}^\dagger \cos \theta + a_m^\dagger i \sin \theta]]] \\[5 pt]
       \nonumber & \text{and so on till}\\[5 pt]
      & a_{m-1}^\dagger \rightarrow a_{m-1}^\dagger \textquotesingle =  a_{m-2}^\dagger  i \sin\theta+\cos\theta [a_{m-1}^\dagger \cos \theta + i \sin \theta [a_m^\dagger]]  \\[5 pt]
&  a_m^\dagger \rightarrow
    a_{m}^\dagger \textquotesingle = a_{m-1}^\dagger i \sin \theta + a_m^\dagger \cos \theta \\[5 pt]
      \end{align*}
 In general, this can be written as
 \begin{align}
 \label{creaion}
    & a_k^\dagger \rightarrow a_k^\dagger \textquotesingle = a_{k-1}^\dagger i \sin \theta + \sum_{p=k}^{m-1} (i \sin \theta)^{p-k} \cos \theta  ^{(2-\delta_{1,k} ) } a_p^\dagger+ (i \sin \theta)^{m-k} \cos \theta ^{(1-\delta _{1,k} ) } a_m^\dagger = \mathcal{A}_k
    \\[5 pt]
\nonumber    &\text{where all the inconsistent terms are set to zero.}
      \end{align}

 For example, in the transformation of $a_1^\dagger$ :  $a_{k-1}^\dagger i \sin \theta$ = $a_{0}^\dagger i \sin \theta$ = 0.

Also, in the transformation of
     $a_{m}^\dagger$ :  $\sum_{i=m}^{m-1} (i \sin \theta)^{i-m} \cos \theta  ^{(2-\delta_{1,m} ) } a_i^\dagger$ =0.

For $m$ ports,
 \begin{align}
 \label{eqn:transformedstate}
U_{m(m-1)}&U_{(m-1)(m-2)}... U_{32}U_{21} \ket{11...\text{m times}.. 11} = \prod_{k=1}^{m}  \mathcal{A}_k
     \end{align}
At this point, it is convenient to define a function $\mathcal{F}$, such that $\mathcal{F}(a_k^{\dagger}$) = sum of coefficients of $a_k^{\dagger}$ in $a_1^{\dagger '}$, $a_2^{\dagger '}$ ... $a_m^{\dagger '}$. Hence we have
\begin{align}
\mathcal{F}(a_k^{\dagger}) = \sum_{j=1}^{k}[(i \sin \theta)^{k-j} (\cos \theta)^{2-\delta(j,1)-\delta(k,m)} ]_j + [(i\sin\theta)^{1-\delta(k,m)}]_{k+1}, \end{align}
where the subscripts $j$ and $k+1$, referred
to as labels, indicate that the corresponding term comes from $a_j^{\dagger}$ and $a_{k+1}^{\dagger}$ respectively.


Now we need to expand the RHS of equation \ref{eqn:transformedstate}. It may seem so that the RHS is constituted of states which are the permutations of 0 and 1 upto $m$ number of terms. But this is not the case and hence, we have to find the allowed states and also their coefficients.

For the first part, we need to find the allowed permutations of $a_1^\dagger$ to $a_m^\dagger$. In the transformed state, a generic $a_k^\dagger$ can at most take power from $0,1,2...(k+1)$, except when $k\neq m$ and upto $m$ when $k=m$. Let a generic  state be $a_1^{\dagger i_1}a_2^{\dagger i_2}...a_m^{\dagger i_m}\ket{000...0}$. So $i_1$ can take values from $0$, $1$, and $2$. After fixing $i_1$, the $i_2$ can take values from zero to min($3,m-i_1$). So, ignoring the coefficients for now, the allowed states after transformation of $\ket{1,1...(m\; \text{times)}}$ are:
\begin{alignat}{3}
& \quad \quad \sum_{i_1=0}^{2}\sum_{i_2=0}^{min(3,m-i_1)} &&...\sum_{i_{m-1}=0}^{min(m,m-\sum_{k=1}^{m-2}i_k)}&&a_1^{\dagger i_1}a_2^{\dagger i_2}...a_m^{\dagger i_m}\ket{000...0} \\ & =&& \Delta_{\text{c}}&& a_1^{\dagger i_1}a_2^{\dagger i_2}...a_m^{\dagger i_m}\ket{000...0}
\end{alignat}
where,
 \begin{itemize}
   \setlength\itemsep{1em}
     \item $\Delta_{\text{c}} \Rightarrow \sum_{i_1=0}^{2}\sum_{i_2=0}^{min(3,m-i_1)} ...\sum_{i_{m-1}=0}^{min(m,m-\sum_{k=1}^{m-2}i_k)} $
    \item $i_m=(m)-\sum_{k=1}^{m-1}i_k$
 \end{itemize}

 Now having found the allowed states, we proceed to find the coefficients of the states. Let us consider the coefficient of the state $a_k^{\dagger i_k}\ket{000.. \text{m times}...0}=\\ \ket{00..\underbrace{i_k}_{\text{$k^{th}$ position}}00.. \text{m times}...0}$. The coefficients of this state can come from exactly $i_k$ out of $a_j^{\dagger}$s, with $0 \leq j \leq k+1$. Let us assume $k=1$. If $i_k$=1, the coefficient comes from either one of $a_1^{\dagger'}$ or $a_2^{\dagger'}$. So the coefficient, including the label as the subscript, is written as [$\cos\theta]_1$+[$i\sin\theta]_2$ = $\mathcal{F}(a_1^{\dagger}$). If $i_k$=2, then the coefficient is $[i\sin\theta]_1 [\cos\theta]_2$ = $\frac{1}{2!}[\mathcal{F}(a_1^{\dagger})]^{2}$, with the added restriction that we ignore the terms with repeated labels.

So the final state after the unitary transformation can be written as
\begin{align}
\label{first}
  U\ket{11...\text{m times}.. 11} =  \Delta_{c} \prod_{p=1}^m \frac{1}{i_p!}[\mathcal{F}(a_p^{\dagger})]^{i_p} a_1^{\dagger i_1}a_2^{\dagger i_2}...a_m^{\dagger i_m}\ket{000...0}
\end{align}

\subsection{Extending to general $m,n$}
Having obtained the output for $m=n$ case, we now focus on the more general case of $m\neq n$. We do so by modifying equation \ref{first} that is, by introducing a way to delete some photons of that input. This can be done using annihilation operators.

As discussed before, the creation operator acts on a state to add a photon to that state, i.e $a^\dagger \ket{0} = \ket{1}$. The inverse of the creation operator is the annihilation operator $a$, such that $a \ket{1} = \ket{0}$.

 Hence,

 \begin{center}
      $a_1\ket{11}$ = $\ket{01}$ $\Rightarrow$ $U_{12}( a_1\ket{11})$ =  $U_{12}\ket{01}$
 \end{center}
      from which we find that $U_{12} a_1 \ket{00} = (a_1 \cos \theta - a_2 i \sin \theta) \ket{00}$. Similar reasoning gives $U_{12} a_2 \ket{00} = (a_2 \cos \theta - a_1 i \sin \theta) \ket{00}$.

 Let us now consider an $m$-channel $n$-input scenario. The input state is $\rho$ as given in equation \ref{eqn:somelabel}. This can be re-written as :
 \begin{align}
 \label{rho}
   \rho =  \sum_{k_1=1}^{m-(m-n)+1} \quad \sum_{\substack{k_2=1\\k_2 > k_1}}^{m-(m-n)+2} ...\sum_{\substack{k_{(m-n)}=1\\k_{(m-n)}> k_1,k_2..k_{(m-n-1)}}}^m \ket{\phi}\bra{\phi}
 \end{align}
with
  \begin{align*}
     \ket{\phi}=a_{k_1} a_{k_2} ... a_{k_{(m-n)}} a_1^\dagger a_2^\dagger ... a_m^\dagger\ket{00..\text{m times}.. 00}.
 \end{align*} where $k_1, k_2$ ... $k_{(m-n)}$ are the channels with zero photons and $k_1<k_2<$...$k_{(m-n)}$.

 This state undergoes transformation as :
 \begin{equation*}
     U_{m(m-1)}U_{(m-1)(m-2)}... U_{32}U_{21} a_{k_1} a_{k_2} ... a_{k_{(m-n)}} a_1^\dagger a_2^\dagger ... a_m^\dagger\ket{00..\text{m times}.. 00}
 \end{equation*}
 The transformation of $a_k^\dagger$ s is already worked out in equation \ref{creaion}. Proceeding in a similar manner for $a_k$'s, we obtain
 \begin{align}
 \label{anihilation}
    a_k \rightarrow a_k'& = a_{k-1} (-i \sin \theta) + \sum_{p=k}^{m-1} (-i \sin \theta)^{p-k} \cos \theta  ^{(2-\delta_{1,k} ) } a_i\\ \nonumber & + (-i \sin \theta)^{m-k} \cos \theta ^{(1-\delta_{1,k} ) } a_m
\end{align}

Consider the products of $a_{k_1}' a_{k_2}' ... a_{k_{(m-n)}}'$. The terms containing $a_{k_1-1}$ to $a_{k_2-2}$ can appear at most once (when $k_1 = 1$, $a_{k_1-1}$ = $a_{0}$, which we assume equals zero), the terms from $a_{k_2-1}$ to $a_{k_3-2}$ at most twice, and so on till $a_{k_{(m-n)}-1}$ to $a_{m}$ appear at most $(m-n)$ times. Hence, for given values of $k_1,k_2...k_{(m-n)}$, all the allowed permutations can be written as :

\begin{alignat}{3}
  &\sum_{i_{(k_1-1)}=0}^{1}\sum_{i_{k_1}=0}^{\mathcal{M}_1(i_{k_1})}... &&\sum_{i_{(k_2-1)}=0}^{\mathcal{M}_2(i_{(k_2-1)})} ...\sum_{i_{(m-1)}=0}^{\mathcal{M}_{m-n}(m-2)}&&a_{k_1-1}^{i_{k_1-1}}a_{k_1}^{i_{k_1-1}}...a_m^{i_m} \ket{000...0}\\
  &=&& \Delta_{a}&& a_{k_1-1}^{i_{k_1-1}}a_{k_1}^{i_{k_1-1}}...a_m^{i_m}
\end{alignat}

 where,
 \begin{itemize}
  \setlength\itemsep{1em}
     \item $\mathcal{M}_x(i_y)=min(x,m-n-\sum_{l=0}^{y-1} i_{l})$
     \item $\Delta_{a} \Rightarrow \sum_{i_{(k_1-1)}=0}^{1}\sum_{i_{k_1}=0}^{\mathcal{M}_1(i_{k_1})}... \sum_{i_{(k_2-1)}=0}^{\mathcal{M}_2(i_{(k_2-1)})} ...\sum_{i_{(m-1)}=0}^{\mathcal{M}_{m-n}(m-2)}$
     \item $i_m=(m-n)-\sum_{j=k_1-1}^{m-1}i_j$
 \end{itemize}
 \smallskip

 We can define $\mathcal{F}$ as we did before, such that:
 \begin{equation}
 \label{F2}
 \mathcal{F}(a_k)=\begin{cases}
    \quad \quad (-i\sin\theta)^{|k-k_1|}(\cos \theta  ^{(2-\delta_{1,k_1} )-2\delta_{k_1-1,k} })_{k_1}, & \text{if $k_1-1\leq k\leq k_2-2$}\\[10pt]
    \sum_{p=1}^2 (-i\sin\theta)^{|k-k_p|}(\cos \theta  ^{(2-\delta_{1,k_p} )-2\delta_{k_p-1,k} })_{k_p},  & \text{if $k_2-1\leq k\leq k_3-2$}\\[10pt]
    \quad \quad \quad \quad \quad\quad\quad\quad\quad\quad\quad\quad\quad\quad \vdots\\[10pt]
        \sum_{p=1}^{m-1} (-i\sin\theta)^{|k-k_p|}(\cos \theta  ^{(2-\delta_{1,k_p} )-2\delta_{k_p-1,k} })_{k_p},  & \text{if $k_{(m-n)}-1\leq k\leq k_m-1$}\\[10pt]
    \sum_{p=1}^{(m-n)} (-i \sin \theta)^{m-k_p} \cos \theta ^{(1-\delta_{1,k_p}} )_{k_p}, & \text{if k=m}
      \end{cases}
\end{equation}

 So the state $\ket{\phi}$ after transformation can be written as
 \begin{align}
  \label{name}
     \ket{\phi} \rightarrow \ket{\phi'} =
        \left(\Delta_{c} \prod_{p=1}^m \frac{1}{i_p!} [\mathcal{F}(a_p^{\dagger})]^{i_p}\right) \left( \Delta_{a} \prod_{j=k1-1}^{m}\frac{1}{i'_j !}[\mathcal{F}(a_{j})]^{i'_j }\right)\ket{\phi}
 \end{align}


  Now, the same has to be done for all $\ket{\phi}$s.
 Using equations \ref{name} and \ref{rho}, the input after transformation can be written as,

\begin{align}
\label{tentative result}
  U\rho = \sum_{k_1=1}^{m-(m-n)+1} \quad \sum_{\substack{k_2=1\\k_2 > k_1}}^{m-(m-n)+2} ...\sum_{\substack{k_{(m-n)}=1\\k_{(m-n)}> k_1,k_2..k_{(m-n-1)}}}^m \ket{\phi'}\bra{\phi'}
\end{align}



   Thus, equation \ref{tentative result} gives transformed state for a general $(m,n)$ input obtained.


\section{Conclusion}

In this paper, we have shown that a fully mixed state of photons can be converted into a maximally entangled states, via an optical network made up of sequential beam splitters. Suitable choices of transmissivity and reflectivity of the beam splitters can maximize the probability of the those entangled states. Our results, though probabilistic, can be useful in quantum communication network based on photons. We further presented a detailed analysis of obtaining the final state of an arbitrary number of photons, initially prepared with a classical probability distribution.

%
%

\begin{acknowledgements}
One of us (H.S.V.) acknowledges financial support through INSPIRE fellowship program, DST, Govt. of India and IIT Ropar for hosting during summer internship program, 2019\end{acknowledgements}

%
%

\bibliographystyle{spphys}       
\bibliography{ref}   

\end{document}